\documentclass[11pt]{article}
\usepackage{jcappub}
\usepackage[utf8x]{inputenc}
\usepackage{epstopdf}

\title{Dark matter Annihilation in the Most Luminous and the Most Massive Ultra-compact Dwarf Galaxies (UCD)}

\author[a,b]{Elaine C. F. S. Fortes}%
\affiliation[a]{Universidade Federal do Pampa\\
	Rua Luiz Joaquim de Sá Brito, s/n, Promorar,\\ Itaqui - RS, 97650-000,
	Brazil}
\affiliation[b]{Instituto Nacional de Pesquisas Espaciais\\
	Av. dos Astronautas, 1758 - Jardim da Granja, \\São Jose dos Campos, SP, 01506-000, Brazil}

\author[b]{Oswaldo D. Miranda}%

\author[c]{Floyd W. Stecker}%
\affiliation[c]{Astrophysics Science Division, \\ NASA Goddard Space Flight Center, Greenbelt, MD 20771
}

\author[b]{Carlos A. Wuensche}%

\emailAdd{elainefortes@unipampa.edu.br}
\emailAdd{oswaldo.miranda@inpe.br}
\emailAdd{floyd.w.stecker@nasa.gov}
\emailAdd{ca.wuensche@inpe.br}

\abstract{In this paper, we explore the potential astrophysical signatures of dark matter (DM)  annihilations in ultra-compact dwarf galaxies (UCDs) considering two of the richest known galaxy clusters within 100 million light-years, nominally, Virgo and Fornax. Fornax UCD3 is the most luminous known UCD and M59 UCD3 is the most  massive known UCD. With the detection of a 3.5 million solar mass black hole (BH) in Fornax UCD3, we carefully model several dark matter (DM) enhanced profile scenarios, considering the presence of both a supermassive black hole (SMBH) and DM. For Fornax UCD3, the comparison of the stellar and dynamical masses suggests that there is little content of DM in UCDs. M59 UCD3 did not receive the same attention in simulations as Fornax UCD3, however deep radio imaging and X-ray observations were performed for M59 UCD3 and can be used to place limits in DM content of these UCDs. We take an average estimate of dark matter content and used the Salpeter and Kroupa mass functions. We model Fornax UCD3 and M59 UCD3 to have a DM content that is the average of these mass functions. We then analyze the constraints for Fornax and M59 UCD3 coming from $\gamma$-ray and radio sources, considering a dark matter particle with a mass between 10 and 34 GeV in our simulations. In the absence of a strong $\gamma$-ray signature, we show that the synchrotron emission from electrons and positrons produced by DM annihilations can be very sensitive to indirect DM search. We find that DM parameters can be significantly constrained at radio frequencies and that the spike profiles play an interesting role in the deep study of the enhancements of DM \& BH interactions in ultra-compact galaxies.}

	
\begin{document}	
\maketitle

\section{Introduction}
\label{sec:intro}

${}$

 Since the discovery of ultra-compact dwarfs galaxies (UCDs), many attempts at observing and important theoretical simulations were performed to shed light into their nature and origin.
 UCDs are, arguably, the  densest stellar systems, brighter and larger than globular clusters,  with $M > 2\times 10^{6}M_{\bigodot}$ and radii $r > 10$ pc which exhibit properties such as mass, luminosity, and size that challenge the conventional understanding of them in comparison to canonical stellar systems.  Since their discovery in 1999 \cite{Drinkwater:2000hr} they have been intensively studied by many astrophysicists who try to find reasonable explanations for their nature, origin, formation mechanisms, and dynamical evolution\cite{Seth}.   Their origin is one of the main open questions in extragalactic astrophysics, leading to many possibilities of interpretations, considering UCDs to be either the result of the evolution of primordial density fluctuations or formed through mergers of globular clusters and considering UCDs to be the nuclei of tidally stripped nucleated dwarf elliptical (dE) galaxies, etc.~\cite{Chilingarian:2008yw, Fellhauer:2001gb,Fellhauer:2005xx, Drinkwater:2003nu,Strader:2013fza}. UCDs were also dubbed as dwarf-globular transition objects (DGTOs) in attempt to express their uncertain origin \cite{Hasegan:2005ct}.

Further evidence suggests that either by considering UCDs to be massive globular clusters \cite{Fellhauer:2001gb,Fellhauer:2005xx} or to be tidally stripped remnants of dwarf galaxies \cite{Drinkwater:2003nu,Strader:2013fza}, both origins could contribute to the observable UCD population \cite{Norris:2011cr}. The simulations presented in ref. \cite{Chilingarian:2008yw} supported the hypothesis of tidal stripping of nucleated dwarf elliptical galaxies and the formation of tidal super-globular clusters in galaxy mergers. Some other alternatives for UCD formation are also discussed in ref. \cite{Chilingarian:2008yw}.

Observations of the stellar orbit velocities in UCDs have suggested that some of them  could host a central supermassive BH, further exciting the search for additional information about UCDs.
The combined observational results of Hubble Space Telescope (HST) and the measurements of integrated velocity dispersion have shown that the dynamical mass to light ratios, i.e., the $M/L$ relations for UCDs were systematically increased if compared to conventional stellar systems \cite{Frank:2011ji,Strader:2013fza}. One of the suggestions to explain the large $M/L$ relation claims for a relic of a massive progenitor galaxy in the tidal stripping scenario, where a central massive black hole (BH) accounts for ~10-15\% of the total mass \cite{Mieske:2013nla}.

In fact, supermassive black holes (SMBH) were confirmed  in four UCDs with $M > 10^{7}M_{\bigodot}$ \cite{Seth}, supporting the idea that UCDs with large $M/L$ host  SMBHs. One of these SMBHs was found in Fornax cluster. The most massive  known UCD M59-UCD3 does not have a confirmed SMBH, but this could be due to the need for better resolution imaging required to model this UCD \cite{Sandoval2015}. However,  M59-UCD3 is an important object for testing the idea that the most massive UCDs host SMBHs. Simulations performed for M59-UCD3 presented an estimate for the BH mass to be in the order of $\sim4.2^{+2.1}_{-1.7}\times 10^{6}M_{\bigodot}$. For Fornax UCD3, the most luminous known UCD, it was detected a  BH with a mass of $3.3^{+1.4}_{-1.2}\times 10^{6}M_{\bigodot}$ in the center of UCD3 galaxy. Both estimates reinforce the hypothesis of tidal stripping \cite{Afanasiev,Seth}.

Stellar population constraints on the DM content in UCDs were discussed in ref. \cite{Chilingarian:2008yw}. Comparing the stellar and dynamical masses, it was suggested that UCDs have little DM content.
Simulations presented in ref. \cite{Chilingarian:2008yw} indicate that low DM content could leave an open door for other UCD formation scenarios \cite{Bekki:2003va}. In the case of dwarf elliptical stripping, the progenitor's nucleus must not be DM dominated. The remaining alternative scenarios as globular cluster merging and formation of UCDs as tidal super-globular clusters assume no DM content. Even taking into account the latter considerations, the large variation of $M/L$ ratio of UCDs might suggest the presence of DM. Thus, UCDs can be considered super-globular clusters (SGC) and there is some evidence of DM presence in GC\cite{Brown:2018pwq}.  Both globular clusters and UCDs could derive from the same formation scenario that involves DM density perturbations. So, the above reasoning is, in itself a good motivation to search for DM signals in UCDs.

In this paper, we will study a possible DM signature in Fornax UCD3 and M59 UCD3, assuming the existence of both DM and a black hole BH in these sources. We will study three types of DM spike density profiles for DM annihilations in these UCD3, considering DM annihilation into leptonic channels and into the $b\bar{b}$ channel. These annihilation channels and the mass range of DM candidates are the ones used to fit the galactic $\gamma-ray$ excess and the galactic isotropic radio emission which is significantly brighter than the expected contributions from extragalactic sources.

The paper is organized as follows: in section \ref{sec:Fornax}, we review some basic information about Fornax UCD3 and M59 UCD3. In section  \ref{sec:Synchrotron} we review the techniques for calculation of synchrotron emission from DM annihilations.  In section \ref{sec:Profiles} we illustrate the DM density ``spike" and ``mini-spike" profiles to be considered here and the parameters set used in our simulations. In section \ref{sec:Numerical} we present the numerical results of the synchrotron flux for Fornax UCD3 and in  section \ref{sec:Conclusion}, we summarize our conclusions about DM presence in this UCD.

\section{Ultra-compact Dwarf Galaxies Fornax UCD3 and M59 UCD3}
\label{sec:Fornax}
${}$

Fornax UCD3, the most luminous known UCD, and M59 UCD3, the most massive known UCD are the subjects of our studies here. Fornax UCD3 is located 11 kpc from the neighboring giant elliptical galaxy NGC 1404 in the central part of Fornax galaxy cluster. Its distance from Earth is assumed to be 20.9 Mpc \cite{Afanasiev}. Recently, a BH with a mass of $3.3^{+1.4}_{-1.2}\times 10^{6}M_{\bigodot}$ was detected in the center of Fornax UCD3 galaxy, corresponding to 4\% of the stellar mass.

M59 UCD3 is located  10.2 kpc in projection from the center of M59 host galaxy, considering an average distance of 16.5 Mpc to Virgo Cluster. Its distance from Earth is 14.9$\pm$ 0.4 Mpc\cite{Blakeslee:2009tc}. The examination of internal properties of M59 UCD3 and deep radio imaging of it were used to estimate an SMBH of mass  $4.2^{+2.1}_{-1.7}\times 10^{6}$ \cite{Seth}.

Besides the supermassive BH, there could be dark matter in these UCDs. In fact, the $M/L$ ratio of UCDs, which vary fairly significantly suggests the presence of DM in some of them. The higher the mass-light ratio ($M / L$) in a system, the stronger the indication of the presence of dark matter in that system. The closer $M / L$ is to the solar proportion, the less weight (or no weight) the dark matter will have. In principle, the increase in the $M / L$ ratio can also be explained by the variation in the typical ages of younger and older populations. However, for systems with high values ($M / L$), the indication and importance of dark matter grow\cite{Bahcall:2013epa, Baumgardt:2008zt}.

Fornax UCD3 was the target of additional simulations like the one presented at refs.\cite{Frank}. Considering integral field spectroscopy data,  they studied the internal dynamics of this dwarf galaxy.  In their study, it was concluded that there could exist DM  content in UCDs.  They had modeled  a DM density profile  expressed by $\rho=\rho_{s}(1+r^{2}/r_{s}^{2})^{-1.25}$ with $r_{s}=200$ pc, which was able to mimic the DM density in Fornax UCD3 \cite{Goerdt:2007rg} and lead to the best fit of DM fraction of $\sim $66\% inside $r < $ 200 pc and a stellar $M/L_{V}=4.3$. In a more detailed work, ref. \cite{Frank:2011ji} calculated the constraints on DM  content for this UCD. They had performed their simulations considering DM fraction of 20\%, 33\%, 60\% and 67\%  of dark matter within a three-dimensional radius of 200 pc, concluding that a DM fraction of 33\% was compatible at the 1$-\sigma$ level with the observed velocity dispersion profile, yielding a stellar $M/L_{V}$ of 3.4. The model with a 60\% of DM fraction was excluded by the data with 98\% of confidence level. With all these simulations performed at the ref. \cite{Frank:2011ji}, they concluded that a significant DM component could strongly suggest UCDs as the remnant nucleus of a larger galaxy or a remnant star cluster around a recoiling supermassive black hole. The internal kinematics of Fornax UCD3 were fully consistent with it being a massive globular cluster.

Other interesting studies presented at the ref. \cite{Chilingarian:2008yw} placed constraints on DM  in UCD galaxies and studied the stellar population parameter in six Fornax clusters UCDs, including UCD3. They were not able to give a conclusive answer if there is DM content in UCDs or not, but they concluded that UCDs are not dominated by DM. This conclusion was based on the models used to derive dynamical and stellar masses. Using the Salpeter and Kroup mass functions and the dynamical mass of the objects, they placed the limits for DM content. Their conclusions are presented in Table 3 of ref.\cite{Chilingarian:2008yw}.

To simulate the DM content in Fornax, we take an average of the 10 predicted estimates of DM content for the five UCDs on that table. We will assume a DM fraction of $\sim$ 8\% and $\sim$ 32\% for Fornax UCD3. The concentration of $\sim$ 32\%  is half the expected value for DM fraction of ref.\cite{Frank} and is close to the fraction claimed to be compatible at 1$-\sigma$ level with the observed velocity dispersion profile \cite{Frank:2011ji}.

M59 UCD3 was not a target of many simulations as were done for Fornax UCD3, but deep radio imaging and X-ray detections of three Virgo cluster were successful in placing upper limits for M59 UCD3 radio flux. In the absence of experimental data for Fornax UCD3, considering similar morphological characteristics of Fornax UCD3 and M59 UCD3 (such as mass, age, size, radius, distance), we can take the derived upper limits to constrain DM content in both UCDs.

In Table \ref{Tab1} we present the main information to be used in our simulations considering the UCD3.

${}$
	\begin{table}[ht]
	\centering
\begin{tabular}{ |p{4.5cm}||p{2.4cm}|p{2.3cm}|  }
 \hline
 \multicolumn{3}{|c|}{Characteristics of UCD Galaxies} \\
 \hline
 & Fornax UCD3 & M59 UCD3 \\
 \hline

 \textit{d} (Mpc) &$20.9\pm 0.3\pm 1.4$ & $14.9\pm 0.4$\\\hline
  $t_{BH}$(Gyr) & $\sim 5$  & $8.6\pm 2.2$  \\\hline
 $M_{BH} (M_{\odot})$ & $3.3_{-1.2}^{+1.4} \times 10^{6}$  & $4.2^{+2.1}_{-1.7}\times 10^{6}$   \\\hline
 $\sigma_{*}$ (km/s) & $33.0 \pm 4.7$  & $77.8 \pm 1.6$ \\\hline
 $R_{a}$ (pc) & 92.2 & $20 \pm 4.2 \pm 0.54$ \\\hline
 $M/L$  & 3.35  & $4.2 \pm 0.4$ \\\hline

  \end{tabular}
    \caption{Data for UCD Fornax UCD3 and M59 UCD3. The parameter  \textit{d} is the distance of these UCD from Earth \cite{Seth, Blakeslee:2009tc, Afanasiev}, $t_{BH}$ denotes the age of the central SMBH in UCD3 \cite{HilkerMichael,Sandoval2015}. We chose the age of the stellar system around the BHs as the age of the black hole ($t_{BH}$). $M_{BH}$ denotes the mass of the SMBH \cite{Afanasiev, Seth} and $\sigma_{*}$ is the stellar velocity dispersion \cite{Afanasiev, ChengzeLiu}, $R_{a}$ is the diameter size of the object \cite{ Hilker:2006di, Sandoval2015} and $M/L$ is the mass-to-light ratio\cite{Afanasiev, Seth}. We emphasize that the central values of these parameters were used in our simulations. Early references considering other numerical values for the parameters can be found in refs.
    \cite{Drinkwater2}, \cite{Chilingarian:2008yw}, and \cite{Hilker:2006di}. }
   \label{Tab1}

  \end{table}

\newpage
\section{Synchrotron/Radio Flux From Dark Matter Annihilation}
\label{sec:Synchrotron}
${}$
A complete treatment of DM  synchrotron emission must consider  the diffusion and energy-loss contribution from secondary particles. Both of these two mechanisms are considered in the diffusion equation, which, after neglecting re-acceleration and convection effects can be expressed as:

\begin{equation}\label{p1}
\dfrac{\partial}{\partial t}\dfrac{dn_{e}}{dE}= \triangledown\left[ D(E,\textbf{r})\triangledown \dfrac{dn_{e}}{dE}\right]\\
+ \dfrac{\partial}{\partial E}\left[ b(E,\textbf{r})\dfrac{dn_{e}}{dE}\right]+Q(E,\textbf{r}),
\end{equation}
where $Q(E,\textbf{r})$ is the source term, $D(E,\textbf{r})$ is the coefficient for spatial diffusion and $b(E,\textbf{r})$ is the loss term.
The source term is expressed as:

\begin{equation}\label{source}
Q(E,r)=\dfrac{\langle\sigma v\rangle \rho(r)^{2}}{2 m_{DM}^{2}}\dfrac{dN}{dE_{inj}},
\end{equation}
where $\langle\sigma v\rangle$ denotes de annihilation cross section, $\rho(r)$ is the spatial distribution of DM, $m_{DM}$ is the mass of DM candidate and $dN/dE^{\prime}_{inj}$ is part of the source term which relates the electron injection spectrum from DM annihilations.

References \cite{Colafrancesco:2006he} and \cite{Colafrancesco:2005ji} had shown that the effect of diffusion on the emitted flux is highly significant in small structures such as dwarf galaxies\cite{Beck:2019ukt}.

The analytical solution for Eq. \ref{p1} for the case of DM source function can be found in ref.\cite{Colafrancesco:2005ji} and numerical treatment for the spatial diffusion coefficient considering specific astrophysical systems can also be found in refs.\cite{Colafrancesco:2006he} and \cite{Baltz:2004bb}. Here, we will treat the diffusion considering a spatially independent form with a power-law energy dependence. We will use the RX-DMFIT tool to perform our calculations \cite{McDaniel:2017ppt}. This tool presents two forms for the diffusion coefficient: one which considers a simplified power law in energy and the other which embodies the degree of uniformity of the magnetic field. The diffusion coefficients are respectively:
\begin{eqnarray}
D(E)=D_{0}E^{\gamma}\\
D(E)=D_{0}\frac{d_{B}^{2/3}}{B_{avg}^{1/3}}E^{\gamma}
\end{eqnarray}
where $D_{0}$ is the diffusion constant, which values are based on studies done in our galaxy and fall in the range $10^{27}-10^{29}$ cm$^2$s$^{-1}$\cite{Blasi:1999rj, Beck:2019imo, Colafrancesco:1998us, Pfrommer:2009hn}. In our simulations we will consider $D_{0}=10^{28}$ cm$^2$s$^{-1}$. The parameter $d_{B}$ is the minimum uniformity scale of the magnetic field. In Ref. \cite{Miranda:1998ne} it is suggested that magnetic fields in the order of $\mu$-Gauss can maintain coherence in scales up to 100 kpc. So, in principle, for UCDs,  the uniformity/coherence scale of the magnetic field could be of the order of the half-light radius of these galaxies.

An analytic solution for Eq. \ref{p1}, was determined for homogenous diffusion using Green function method \cite{Colafrancesco:2005ji}. Following the notation presented in Ref. \cite{Colafrancesco:2005ji}, the solution of  Eq. \ref{p1} has the form:

\begin{eqnarray}
\dfrac{\partial n_{e}}{\partial E}=\frac{1}{b(E,r)}\int_{E}^{m_{DM}}dE^{\prime}G(r,v(E)-v(E^{\prime}))Q(E,r),
\label{G1}
\end{eqnarray}
where  the Green's function  $G(r,v(E)-v(E^{\prime}))$ is expressed as:
\begin{eqnarray}
G(r,\Delta v)=&&\frac{1}{\sqrt{4\pi\Delta v}}\sum_{n=-\infty}^{\infty}(-1)^{n}\int_{0}^{r_{h}}dr^{\prime}\dfrac{r^{\prime}}{r_{n}}\left(\frac{\rho(r^{\prime})}{\rho(r)}\right)^{2}\\\nonumber
&&\times\left[\exp\left(-\frac{(r^{\prime}-r_{n})^2}{4\Delta v} \right) -\exp\left(-\frac{(r^{\prime}+r_{n})^2}{4\Delta v} \right)  \right].
\end{eqnarray}
For evaluation of the Green's function, free escape boundary conditions at the radius of the diffusion zone ($r_{h}$) are imposed. Besides, the method of image charge is adopted with the charges being placed at $r_{n}=(-1)^{n}r+2nr_{h}$. All information regarding to diffusion coefficient and energy loss terms are incorporated into $\Delta v=v(E)-v(E^{\prime})$. In this case $v(E)$ is defined as:	
\begin{eqnarray}
v(E) = \int_{E}^{m_{DM}}dE^{\prime\prime}\frac{D(E^{\prime\prime})}{b(E^{\prime\prime})}.
\end{eqnarray}
where $\sqrt{\Delta v}$	translates to the mean distance traveled by an electron as it loses energy between its source energy $E^{\prime}$ and the interaction energy, E. An additional information  regarding to the derivation the Green's function according to the method of Ref. \cite{Colafrancesco:2005ji}, is that a spatially independent magnetic field is needed. So, RX-DMFIT tool considers $b(E,r)\approx b(E)$ only for the evaluation of the Green's function\footnote{For the energy loss term outside the integral of Eq. \ref{G1}, the full spatial profile of magnetic field is used. }.

On the other hand, if the diffusion term is neglected, which means that electrons and positrons lose energy on a timescale much shorter compared to the timescale for spatial diffusion, the expression for electron equilibrium spectrum becomes \cite{Storm:2016bfw}:

 \begin{equation}\label{Psyn2}
\dfrac{dn_{e}}{dE}(E,r)=\dfrac{ \langle\sigma v\rangle \rho(r)^{2}}{2 m_{DM}^{2}b(E,r)}\int_{E}^{m_{DM}}dE^{\prime}\dfrac{dN}{dE^{\prime}_{inj}},
\end{equation}
 In our simulations we had worked considering a scenario with diffusion and a scenario with no diffusion. In the next section we will present the DM density profiles to be modelled. For the scenario with no diffusion, which contemplates the profiles Frank, NFW-Spike and spike, we  took $dN/dE^{\prime}_{inj}$ from ref. \cite{Buch:2015iya}. For the scenario with takes diffusion into account and considers NFW DM density profile, we  used RX-DMFIT tool, with obtains $dN/dE^{\prime}_{inj}$ from the package presented in Ref. \cite{Gondolo:2004sc}.

The energy loss term ($b(E,r)$) in Eq. (\ref{Psyn2}) is described by
\begin{equation}
 b(E,r)=b_{syn}+b_{IC}+b_{brem}+b_{coul}
\end{equation}
where the terms $b_{syn}$, $b_{IC}$, $b_{brem}$ and $b_{coul}$ denote the loss by synchrotron radiation, inverse Compton scattering, bremsstrahlung and Coulomb interactions, respectively~\cite{st69}.

In the energy scales that we are considering (10-34 GeV), the DM-induced emission is dominated by the synchrotron radiation and inverse Compton scattering of the relativistic secondary electrons and positrons.

The energy loss due to all these processes can be expressed by \cite{Storm:2016bfw}.

\begin{eqnarray}
b(E,r)&& \approx  0.0254\left(\dfrac{E}{1 GeV}\right)^{2}\left(\dfrac{B(r)}{1\mu G}\right)^{2}+ 0.25\left( \dfrac{E}{GeV}\right)^{2}\\
&&+ 1.51n(0.36+log(\gamma/n))+ 6.13n(1+log(\gamma/n)/75.0),
\end{eqnarray}
where this loss term has units of $10^{-16}$ GeV/s, $E=\gamma m_{e}c^{2}$, $B$ denotes the magnetic field, $n$ denotes the number of free electrons which corresponds to the density of ionized gas.

The synchrotron power radiated by a single electron as a function of frequency is given as follows:

\begin{equation}\label{Psyn}
 P_{\nu}(\nu,E,\alpha)=\dfrac{\sqrt{3}e^{3}B \sin \alpha F(x)}{m_{e}c^{2}},
\end{equation}
with $x=\nu/\nu^{\prime}$, $\nu^{\prime}=\nu_{c}\sin \alpha /2$ and $\nu_{c}=3 e B \gamma^{2}/(2\pi m_{e}c)$,  $e$ denotes the electric charge, $B$ is the magnetic field, $c$ is the velocity of light, $\gamma$ is the Lorentz factor, related to the energy of a single electron by $E=\gamma m_{e}c^2$ and $F(x)$ is given by
\begin{equation}\label{Fx}
F(x)=x \int_{x}^{\infty} K_{5/3}(x^{\prime})dx^{\prime},
\end{equation}
where $K_{n}$  is the modified  Bessel function of the second kind and order $n$.  Averaging  randomly oriented magnetic field over the pitch angle $\alpha$, we can express  $P_{\nu}(\nu,E)$ as:
\begin{equation}\label{Psyn1}
 P_{\nu}(\nu,E)=\dfrac{1}{2} \int_{0}^{\pi} d\alpha \sin \alpha P\nu(\nu,E,\alpha),
\end{equation}
so that we have
\begin{equation}\label{Psyn2}
 P_{\nu}(\nu,E)=2\sqrt{3}\dfrac{e^{3}B}{m_{e}c^{2}}y^{2}\left[K_{4/3}(y)K_{1/3}(y)-\dfrac{3}{5} y \left(K_{4/3}^{2}(y)-K_{1/3}^{2}(y) \right)\right],
\end{equation}
where $y=\nu/\nu_{c}$.

The synchrotron emissivity is related tothe synchrotron power $P_{\nu}(\nu,E)$ and the electron equilibrium energy spectrum $ \dfrac{dn_{e}}{dE}(E,r)$ by

\begin{equation}\label{jnu}
 j_{\nu}(\nu,r)=2 \int_{m_{e}c^{2}}^{m_{DM}}dE \dfrac{dn_{e}}{dE}(E,r)P_{\nu}(\nu,E)
\end{equation}
where $j_{\nu}$ denotes the synchrotron emissivity and the factor 2 takes into account the electrons and positrons.

Finally, the integrated synchrotron flux density produced by a generic distribution reads:
\begin{equation}\label{Fluxo}
 S_{\nu}(\nu)\approx \dfrac{1}{D^{2}}\int_{0}^{R}dr r^{2}j_{\nu}(\nu, r),
\end{equation}
where this result is usually given in Janskys, $R$ is the diameter size of the object of interest. The parameter $D$ is the proper distance.

\section{DM Density Profiles}
\label{sec:Profiles}
${}$

In the simulations performed here for  Fornax UCD3 and M59 UCD3, we consider that an SMBH and DM content cohabit together. SMBHs modify the distribution of dark matter in its vicinity. So, we are working in potential locations for detecting DM indirectly.  DM annihilation/decay products could be searched for in the vicinities of BH since these massive objects could induce overdensities, called ``spikes" or lighter overdensities called ``mini-spikes".

The literature presents many cases of modified profiles, whose authors studied the distribution of DM in the surroundings of a BH. Gondolo and Silk studied DM spike density profile in the galactic center, where a supermassive black hole $2-4\times 10^{6} M_{\odot}$ exist\cite{Gondolo:1999gy, Gondolo:1999ef, Gondolo:2000pn}. Lacroix \textit{et al}, in an alternative way,  modified NFW canonical profile, trying to prove DM density spike allied to the radial dependence at Galactic center\cite{Lacroix:2013qka}. In addition to spike profiles, lighter overdensities around intermediate-mass black holes (IMBH) were analyzed for the first time by Silk and Zhao, who presented the alternative mini-spikes profiles for the IMBH in the Milk Way \cite{Zhao:2005zr}.

In fact, the DM density profile around BHs is still a matter of discussion, despite many suggestions of strong enhancements for it. Certainly, all the range of dynamics effects of BH+DM interaction such as black hole growth time scale, core relaxation time by stellar dynamical heating, the adiabatic response of DM  was not completely mapped\cite{Vasiliev:2008uz, Ullio:2001fb}.

In the simulations performed here, we have considered three types of enhanced DM density profiles, besides the Navarro-Frenk-White. They differ among themselves in many ways, like in the radial dependence, shape, slope of the curve, cusp, spike,  etc.

We will use these profiles to study DM halo in the surroundings of an SMBH in UCD3. We emphasize that the full range of dynamical effects of a BH was not explored in considering the spikes. We just focus on some well-motivated dense inner spike profiles and evaluate their effects on radio signals of Fornax UCD3.

We then compare our simulations with the existing radio flux experimental data.  The considered DM density profiles are respectively

\begin{itemize}
  \item NFW;
  \item NFW + spike;
   \item Mini-spike;
  \item Spike;
\end{itemize}

\subsection{Navarro-Frenk-White  Profile (NFW)}
${}$

The Navarro Frenk-White density profile is given by
\begin{eqnarray}
\rho_{NFW}=\rho_{s}\dfrac{r_{s}}{\, {\rm r}}\left(1+ \dfrac{\, {\rm r}}{r_{s}}\right)^{-2}  & \hbox{} & (r\leq r_{vir})
\end{eqnarray}
where $\rho_{s}$ is the characteristic inner density and $r_{s}$ is the scale radius. These parameters are very sensitive to the epoch of halo formation and correlate very strongly  with the halo virial parameters, via the concentration parameter $c$ and $\delta_{c}$, both dimensionless parameters. The parameter $c=r_{vir}/r_{s}$ and  $\delta_{c}$ are related by
\begin{eqnarray}
\delta_{c}=\dfrac{200}{3}\dfrac{c^{3}}{\ln(1+c)-c/(1+c)},
\end{eqnarray}
where $r_{vir}$ denotes the virial radius. The parameters $c$ and $\delta_{c}$ are also linked by requiring that the mean density at $r_{200}$  be 200$\times \rho_{crit}$, where $\rho_{crit}=3H^{2}/8\pi G=5.62\times 10^{-6}$ GeV/cm$^3$, so that $\rho_{s}=\rho_{crit}\times \delta_{c}$.

The DM content in Fornax UCD3 was simulated at refs.\cite{Frank, Frank:2011ji}, taking $r_{s}=200$ pc. For  galaxies with  different morphological types and a broad range of stellar mass, there is an approximately linear relation between the half-mass radius ($r_{h}$) and the virial radius ($r_{vir}$), $r_{h}\sim 0.015$ $r_{vir}$ \cite{Kravtsov:2012jn}.  Taking 92.2 pc\cite{ Hilker:2006di}  as the half-light-radius of Fornax UCD3 and considering $r_{h} \sim 2\%$ $ r_{vir}$, we can estimate in a conservative way, that the virial radius of Fornax UCD3  is $r_{vir}=4610$ pc, which results in a concentration parameter  $c=23.05$ and consequently $\rho_{s}\simeq 2.07$ GeV/cm$^3$. Ref \cite{Goerdt:2007rg} simulated the formation of UCDs considering two scenarios. In their second simulated scenario, they considered that UCDs are remnants of massive stripped nucleated galaxies. They had set up a spherical equilibrium NFW with parameters compatible with $\Lambda$CDM model and considered a concentration parameter $c=6.2$, concluding that such scenario could enhance a central DM content so that the nucleus could contain a large fraction of DM. Considering the latter value for the concentration parameter as an alternative one for the simulations,  would lead to $\rho_{s}\simeq 0.08$ GeV/cm$^3$. Here we will work with both concentration parameters in order to perform our simulations.
We stress that a DM density profile $\rho=\rho_{s}(1+r^{2}/r_{s}^{2})^{-1.25}$ was the one that mimic DM density in UCDs \cite{Frank,Goerdt:2007rg}.
As previously mentioned, due to similar morphological characteristics between Fornax UCD3 and M59 UCD3, in the absence of parameters for the latter, we will model it, considering the concentration and density parameters inherent to Fornax UCD3.

\subsection{NFW $+$ Spike Profile}
${}$
For the  NFW $+$ spike profile we assume a radial dependence based on the DM density profile presented at \cite{Lacroix:2013qka}.  Their original profile was built to study the strong enhancement due to DM around the SMBH Sgr A$^*$ at the Galactic Center.  Here, we had adapted this profile to study the SMBH in Fornax UCD3 and M59 UCD3, considering typical $\rho_{s}$ and $r_{s}$ of this Fornax UCD3.

This profile is represented by
\begin{equation}
\hbox{$\rho$}(r)=\left\{\begin{array}{lll}  \rho_{s}\dfrac{r_{s}}{r}\left(1+r/r_{s} \right)^{-2} & \hbox{} & r> R_{sp} \\ \rho_{sat}\left(\dfrac{r}{r_{sat}}\right)^{-\gamma_{sp}} & \hbox{} & r_{sat}< r \leqslant R_{sp} \\\rho_{sat} & \hbox{} &  r \leqslant r_{sat} \end{array}\right.
\end{equation}
where,   $r_s =200$ pc parameterizes the NFW profile for Fornax UCD3, described before \cite{Frank}. Here $R_{sp} = G M_{BH}/\sigma_{*}^2$ \cite{Peebles} denotes the radius of the spike, where $G$ denotes the Newton's constant of gravity, $M_{BH}$ is the mass of the black hole and $\sigma_{*}$ is the stellar velocity dispersion. The parameter $\gamma_{sp}=(9-2\gamma)/(4-\gamma)$ and is expected to be between 2.25 and 2.5, taking into account that $0 <\gamma < 2$. In this work we had considered $\gamma_{sp}=2.3$. The parameter $\rho_{sat}$ is the saturation density established by DM annihilations. In this case, this corresponds to make $\rho_{sat}=\rho_{sat}^{ann}$, where
\begin{equation}
\rho_{sat}^{ann}=\dfrac{m_{DM}}{\langle\sigma v\rangle t_{i}},
\end{equation}
basically establishing $\rho_{sat}=\rho_{sat}^{ann}$ corresponds to the equality of characteristic annihilation time and infall time ($ t_{i}$) of DM towards the SMBH. In all the analysed spike profiles, we will assume conservatively that $t_{i}=t_{BH}$, where $t_{BH}$ denotes the age of the black hole, $m_{DM}$ is the mass of DM candidate and $\langle\sigma v\rangle$ is the annihilation cross section. The saturation radius is obtained by requiring that $\rho_{sat}=\rho(r_{sat})$ \cite{Lacroix:2013qka}, so that for Fornax UCD3 we have
\begin{equation}
r_{sat}=R_{sp}\left[\dfrac{\rho_{s}}{\rho_{sat}}\dfrac{r_{s}}{R_{sp}}  \right]^{1/\gamma_{sp}}
\end{equation}

\subsection{Mini-Spike Profile}
${}$
For illustration, we present here the mini-spike profile presented at the ref. \cite{Lacroix:2017uqn}, which was motivated in an attempt to explain the diffuse Fermi-LAT ``excess". In that reference,  they had presented numerical simulations to illustrate how the BH-spike could reproduce the spectrum, profile, and morphology of the Fermi Galactic Center excess\cite{Goodenough:2009gk}, considering the benchmark scenario where DM candidates with a mass $\sim$ 30 GeV annihilates in $b\bar{b}$.

This profile is represented by
\begin{equation}
\hbox{$\rho$}(r)=\left\{\begin{array}{lll}  0 & \hbox{} & r\leqslant 2R_{s} \\\rho_{sat} & \hbox{} & 2R_{s} < r \leqslant R_{sat} \\\rho_{0}\left(\dfrac{r}{R_{sp}}\right)^{-\gamma_{sp}} & \hbox{} &  R_{sat} < r \leqslant R_{sp}
\end{array}\right.
\end{equation}
where $R_{s}=2 G M_{BH}/c^{2}$ denotes the Schwarzschild radius,  $R_{sat}$ is different of $r_{sat}$, presented in the last  profile NFW $+$ spike, $R_{sat}=R_{sp}(\rho_{sat}/\rho_{0})^{-1/\gamma_{sp}}$. In the original reference, the parameter $\rho_{0}$ was obtained by requiring that all the mass inside the spike $M_{sp}$ be of the order of $M_{BH}$ \cite{Lacroix:2017uqn}. In our simulations, as given in section \ref{sec:Fornax}, we required that all the mass inside the spike ($M_{sp}$) be $M_{sp}\approx 8M_{BH}$, which corresponds to $\simeq$ 32\% of the total mass of Fornax UCD3 and also that  all the mass inside the spike be $M_{sp}\approx 2M_{BH}$, which corresponds to $\simeq$ 8\% of the total mass of Fornax UCD3. So, in our approximation   $\rho_{0}\approx (3-\gamma_{sp})2 M_{BH}/(\pi R_{sp}^{3})$ and $\rho_{0}\approx (3-\gamma_{sp})M_{BH}/(2 \pi R_{sp}^{3})$ respectively, considering both scenarios.

\subsection{Spike Profile}
${}$
For the spike profile, we have considered the DM density profile presented at ref.\cite{Brown:2018pwq} which accounts for both the presence of IMBH and dynamical processes in the globular cluster. The spike structure of this profile was motivated by the strong evidence of DM mass with 47 Tuc. They had analyzed the whole set of observations of the 9 years of Fermi-LAT operation in an attempt to find a possible explanation of the $\gamma$-rays excess, not considering only DM but also millisecond pulsars. The spike, in its turn, enhance the $\gamma$-ray signal from DM annihilation.

This profile is represented by
\begin{equation}
\hbox{$\rho$}(r)=\left\{\begin{array}{lll}  0 & \hbox{} & r< 2R_{s} \\\dfrac{\rho_{sp}(r)\rho_{sat}}{\rho_{sp}(r) + \rho_{sat}} & \hbox{} & 2R_{s}\leqslant r < R_{sp} \\\rho_{0}\left(\dfrac{r}{R_{sp}}\right)^{-5} & \hbox{} &  r\geqslant R_{sp} \end{array}\right.
\end{equation}
where $\rho_{sp}(r)=\rho_{0}(r/R_{sp})^{-3/2}$  and in the region outside of the spike, it was assumed that $r^{-5}$ (from tidal stripping) in order to keep few DM content out of cluster \cite{Brown:2018pwq}. The parameters $R_{s}$ and $\rho_{0}$ were defined in the mini-spike profile.  Again, we stress here that we required that all the mass inside the spike  $M_{sp}$ be $M_{sp}\approx 8M_{BH}$ and $M_{sp}\approx 2M_{BH}$, which corresponds to $\simeq$ 32\% and $\simeq$ 8\% respectively of the total mass of Fornax UCD3.

This profile was already used to study a DM population clustered in the vicinities of an SMBH at the center of Centaurus A\cite{Brown:2018pwq, Brown:2016sbl}.

\subsection{A Comparison Between Spiky Profiles}

\begin{figure*}[ht!]
	\centering
	\begin{tabular}{cc}
		\includegraphics[width=0.5\linewidth]{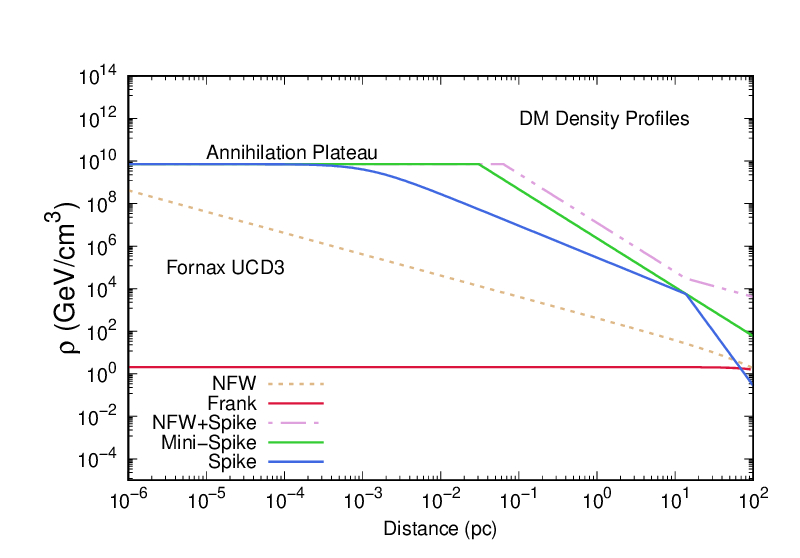}
		\includegraphics[width=0.5\linewidth]{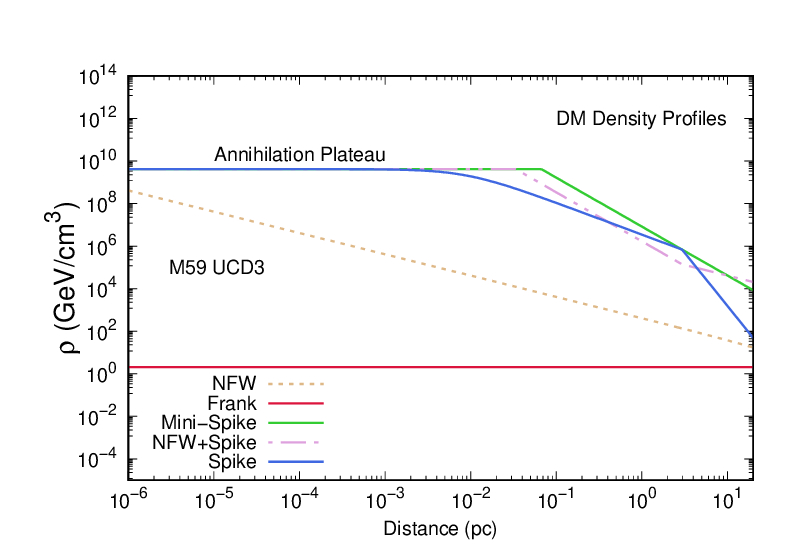}\\
	\end{tabular}
	\caption{The panels show the Spike density profiles created around the SMBH, the canonical NFW profile and the Frank profile predicted in the simulations for Fornax UCD3. We present here three spike models, shown with the colors green (mini-spike), blue (spike), pink (NFW+spike) and the profiles represented by the color orange (NFW) and red (Frank).
		The spikes are formed within the radius of influence of the SMBH of the UCDs. Fornax UCD3 is represented in the left panel, M59 UCD3 is represented in the right panel. The annihilation plateau,  $\rho_{sat}= m_{DM}/(\langle\sigma v\rangle \cdot t_{BH})$ was built considering $m_{DM}=34$ GeV, $\langle\sigma v\rangle=3\times 10^{-26}$ cm$^{3}$/s and $t_{BH}\approx 5$ Gyr for Fornax UCD3 and $t_{BH}\approx 8.6$ Gyr for M59 UCD3.}
	\label{fig1}
\end{figure*}

${}$
In the Figure \ref{fig1} we present an estimate of all these profiles considering, for example, a DM particle of $m_{DM}=34$ GeV and a thermal cross-section $\sigma v=3\times 10^{-26}$ cm$^3$/s. All the remaining data used to build these profiles were taken from Table \ref{Tab1}.

In Figure \ref{fig1}, we show the  DM density profiles considering three kinds of spikes plus Frank and NFW profiles. The later one is represented in dashed orange. The enhancement of the NFW density profile due to an SMBH at the center of the DM system is represented in dashed-dotted pink. In this case, a physical interpretation suggests that DM is gravitationally forced to go near the BH. This situation creates a high concentration of DM, which we call a spike. The spikes are formed with the typical radius of influence of the black hole, $Rsp=G M_{BH}/\sigma_{*}^2$. Comparing NFW, NFW+Spike, Mini-Spike (green), and Spike (blue) profiles we can distinguish 1 cusp \& 3 spike profiles.  The  NFW cusp profile is shallow if compared to NFW+Spike, mini-spike, and spike profiles, which are steeper. This can be explained by the power-law in the density profile $\rho \propto r^{-\gamma_{sp}}$. For NFW+Spike and mini-spike profiles, $\rho \propto r^{-2.3}$ and for the spike profile $\rho \propto r^{-5}$. The constant upper straight line in the graphs indicates that DM density becomes so high in the inner regions of the UCD galaxies holding the SMBH, that annihilations deplete the number of DM particles and this situation produces an annihilation plateau.  Regarding the Frank profile (red), although the density calculated for it appears to be a constant on the graph, it is not. What happens is that the density variations for Frank profile are much smaller compared to the others presented here.

The annihilations of DM become enhanced in spike profiles taking into account a dependence of annihilation rate on $\rho^2$(\textbf{r}), see Eq. \ref{source}. In this way, the study of DM annihilation with enhanced profiles is extremely important for the radio and $\gamma$-rays flux, among others. The spike profiles increase the predicted flux if compared to the canonical non-spiky profiles. The differences that appear in the curves of Fornax UCD3 and M59 UCD3 are due to their intrinsic parameters, such as black hole mass ($M_{BH}$), distance ($d$) from Earth and the stellar velocity dispersion ($\sigma_{*}$), and black hole age $t_{BH}$, all presented in Table \ref{Tab1}.

\newpage
\section{Numerical Results}
\label{sec:Numerical}
${}$

In this section, we present our numerical results for the predicted synchrotron flux considering Fornax UCD3 and M59 UCD3 with the studied profiles. We had included the effects of diffusion in the simulations considering for while only the non-spike density profile Navarro-Frenk-White.

For the diffusion calculations, we adopted a spatially independent form with a power-law energy dependence. We leave the study of effects of diffusion for spike DM density profiles around BHs for a forthcoming work. On it, besides studying diffusion in the region around the BH, we  intend to study some of the strong enhancements of such profiles considering effects of dynamical relaxation by stellar interactions, the loss of DM falling into the BH, and other dynamical effects of BH-DM interactions.

In our work we consider that the fluxes produced by Fornax UCD3 and M59 UCD3 are due to DM annihilation in two hypothetical scenarios $DM,DM \rightarrow b\bar{b}$ and $DM,DM \rightarrow \mu^{+}\mu^{-}$. In our analysis, for the magnetic configurations, we incorporate the observations of the Virgo cluster magnetic field. For typical conditions in the intracluster medium of $n\sim10^{−4}$ cm$^{−3}$ and galaxy velocities $v_{gal}\sim 1,000$ km/s, the maximum magnetic field would be 7$\mu$G \cite{Pfrommer:2009hn}. We have used this value for both Fornax UCD3 and M59 UCD3  considering once again the morphological similarity between these UCDs. Besides, using 7$\mu$G for the magnetic field, for comparison, we chose two more values of the magnetic fields being simulated. They are $B = 1\mu$G and $B = 50\mu$G.

In order to constrain the synchrotron signal from DM annihilation in these UCDs we need to estimate the sensitivity of the existing radio surveys to diffuse emission from these UCDs.  For three UCDs in Virgo Cluster, deep radio imaging data were obtained using the Karl G. Jansky Very Large Telescope Array (VLA)\cite{Seth} for a total of 3.5 hrs of observation. A faint radio emission was detected in M59cO, but only an upper limit of 7.8 $\mu$ Jy at 5.8 Ghz was found for M59 UCD3. We work with a value of 7.8 $\mu$ Jy to model our scenarios.

\begin{figure*}[ht!]
	\centering
	\begin{tabular}{cc}
		\includegraphics[width=0.5\linewidth]{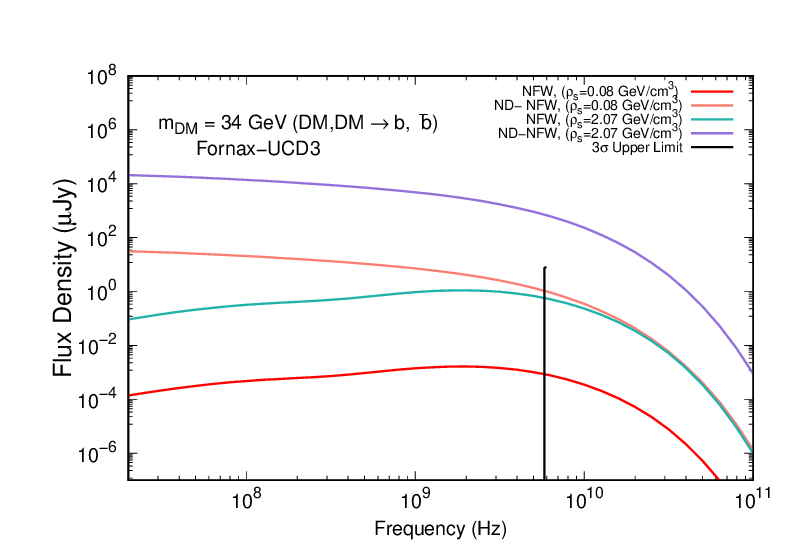}
		\includegraphics[width=0.5\linewidth]{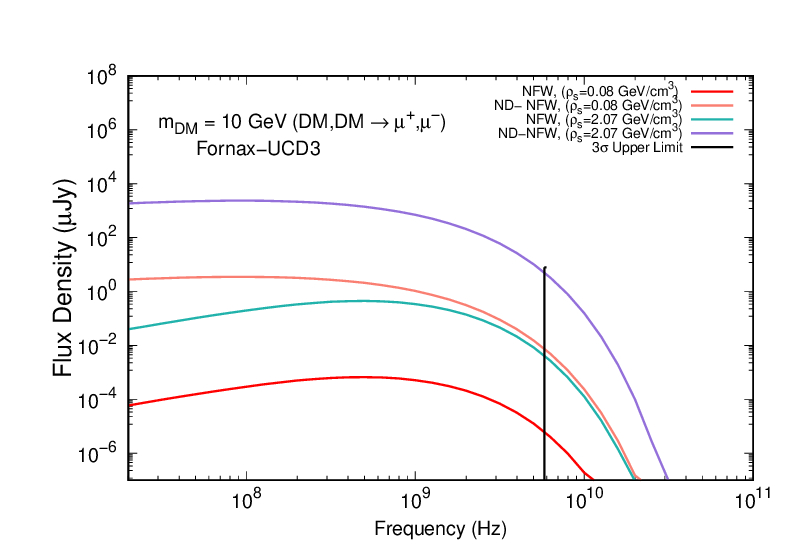}\\
		\includegraphics[width=0.5\linewidth]{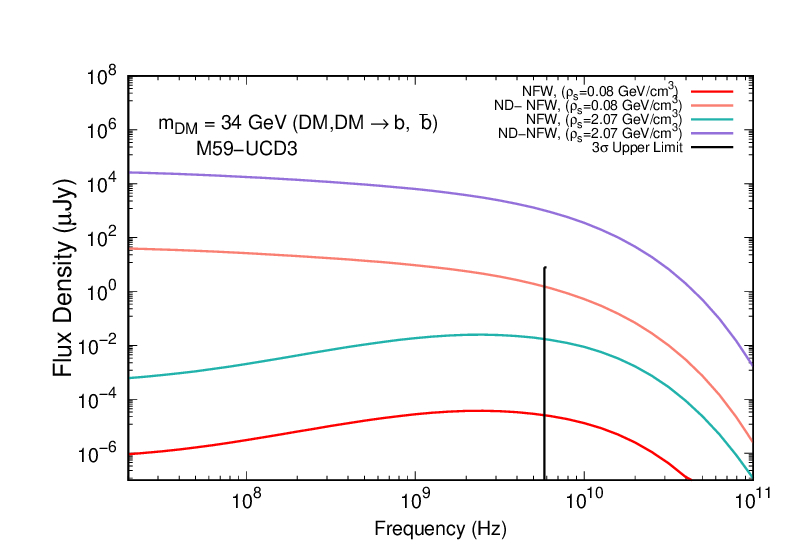}
		\includegraphics[width=0.5\linewidth]{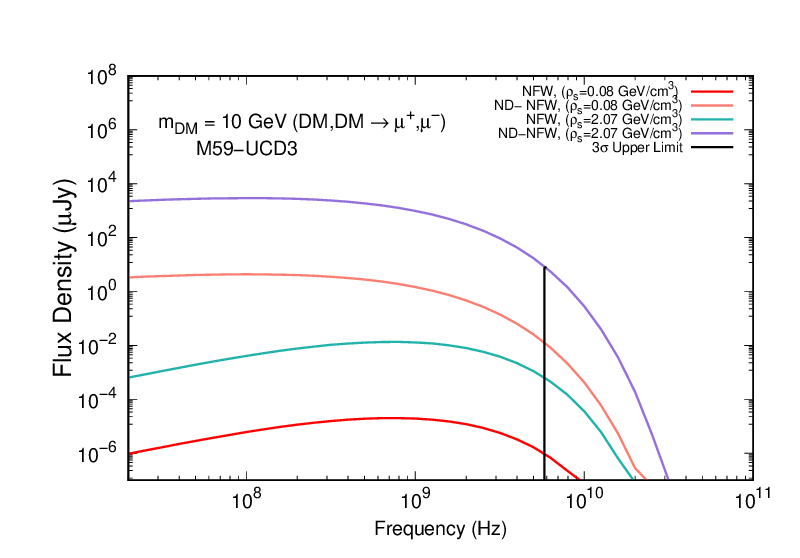}\\
		\end{tabular}
	\caption{The expected flux from DM annihilation versus frequency for Fornax UCD3 and M59 UCD3 considering the NFW DM density profile. In the left panels, we show the synchrotron fluxes  considering $DM,DM \rightarrow b\bar{b}$.  In the right panels, we show the synchrotron fluxes considering $DM,DM \rightarrow \mu^{+}\mu^{-}$. Our DM candidate has mass $m_{DM}=34$ GeV and annihilates in $DM,DM \rightarrow b\bar{b}$ or has mass $m_{DM}=10$ GeV and annihilates in $DM,DM \rightarrow \mu^{+}\mu^{-}$. M59 UCD3  has an experimental upper limit taken from deep radio imaging observations at 5.8 GHz \cite{Seth}, which is represented by the black straight line. The upper limit at that frequency is  7.8 $\mu$Jy. All these fluxes were built considering NFW DM density profile with different DM density. Here the effects of diffusion are included. The index ND means ``no diffusion". }
	\label{fig4}
\end{figure*}

In Figure \ref{fig4}, we show the synchrotron flux for Fornax UCD3 and M59 UCD3 considering the profile NFW.  In these graphs, we had included the effects of diffusion and compared them to the curves where these effects are neglected.  We had also included the upper limit obtained by the radio surveys for M59 UCD3. For these graphs, we had considered $B=7\mu$G and the annihilation cross-section  are $\langle\sigma v\rangle=3\times 10^{-26}$ cm$^{3}/$s for $DM,DM \rightarrow b\bar{b}$ and $\langle\sigma v\rangle=3\times 10^{-28}$ cm$^{3}/$s for $DM,DM \rightarrow \mu^{+}\mu^{-}$. The parameters of the UCDs taken to build the NFW density profile are described in section \ref{sec:Profiles}. In fact, as already argued in section \ref{sec:Synchrotron}, the effect of diffusion is quite significant for compact objects as UCDs. Considering the $DM,DM \rightarrow b\bar{b}$ channel, for Fornax UCD3,  the effect of diffusion decreases the synchrotron flux by $\sim$ 5-6 orders of magnitude as compared to the curves with no diffusion effects (ND). For M59 UCD3, the effect of diffusion decreases by $\sim$7-8 orders of magnitude the synchrotron flux in performing the same comparison. Considering the other channel $DM,DM \rightarrow \mu^{+}\mu^{-}$, for Fornax UCD3, the diffusion decreases the synchrotron flux by $\sim$4-6 orders of magnitude as compared to the curves with (ND) effects. For M59 UCD3, considering the same channel, diffusion decreases the synchrotron flux by $\sim$6-8 orders of magnitude as compared to the scenario where diffusion is not taken into account.

\begin{figure*}[ht!]
\centering
\begin{tabular}{cc}
\includegraphics[width=0.5\linewidth]{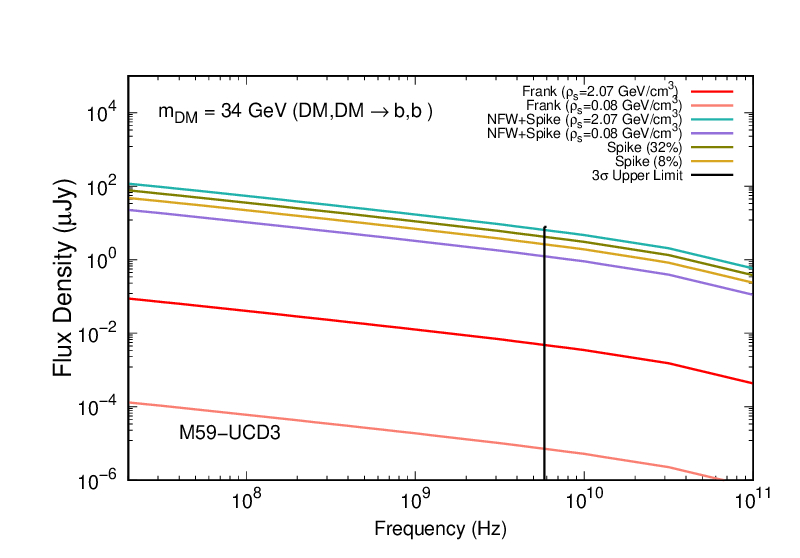}
\includegraphics[width=0.5\linewidth]{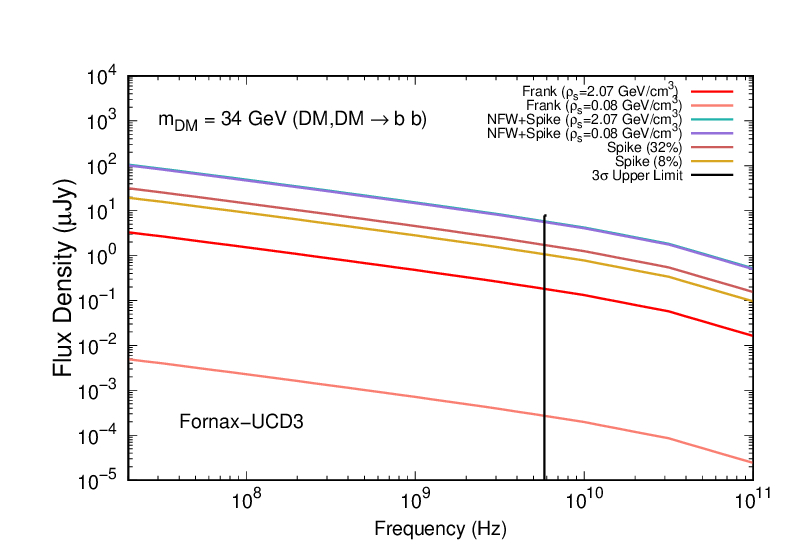}\\
\includegraphics[width=0.5\linewidth]{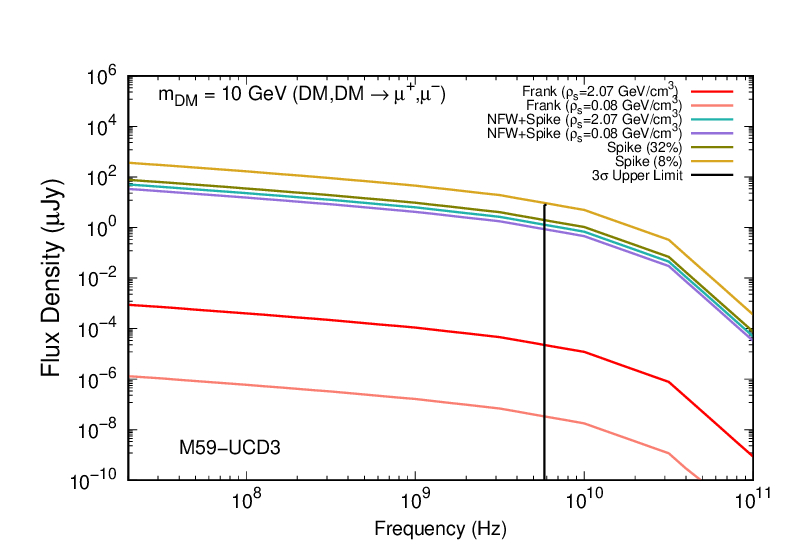}
\includegraphics[width=0.5\linewidth]{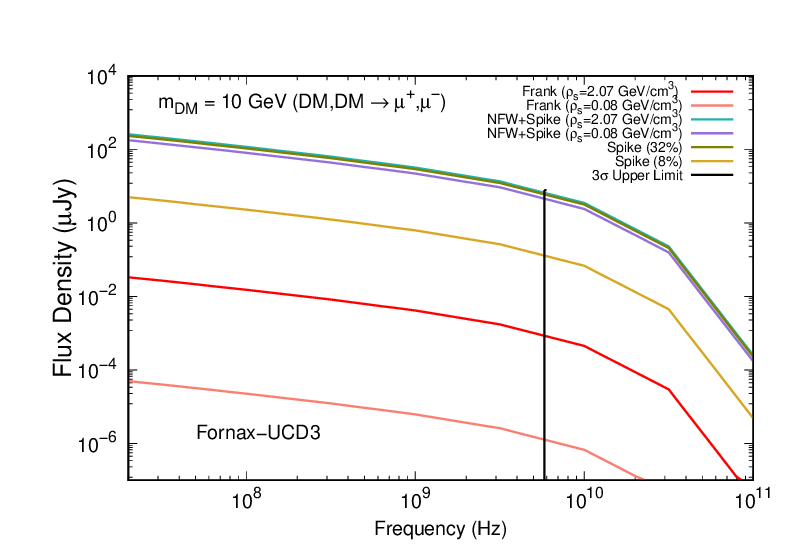}\\
\end{tabular}
\caption{The expected flux from DM annihilation as a function of frequency for Fornax UCD3 and M59 UCD3. In the left panels, we show the synchrotron fluxes results for M59 UCD3 and in the right panels, we present the results for  Fornax UCD3.  Our DM candidate has mass $m_{DM}=34$ GeV and annihilates in $b\bar{b}$ (upper panels) and has mass $m_{DM}=10$ GeV and annihilates in $\mu^{+}\mu^{-}$ (lower panels). All the values used for the annihilation cross-section can be found in Table \ref{Tab2}. M59 UCD3  has an experimental upper limit taken from deep radio imaging observations at 5.8 GHz \cite{Seth}, which is represented by the black straight line. The upper limit at that frequency is  7.8 $\mu$Jy. All these fluxes were built considering Frank, NFW+Spike, and Spike profiles with different DM density and DM concentration.}
\label{fig2}
\end{figure*}

In Figures \ref{fig2}, we show the synchrotron flux for Fornax UCD3 and M59 UCD3 considering the profiles NFW+Spike and Spike. For comparison purposes, we had included the profile used to mimic DM in UCDs dubbed by us as ``Frank" profile. In these graphs, we had also included the upper limit obtained by the radio surveys for M59 UCD3 and they are build with $B=7\mu$G. We can distinguish different density profiles using the synchrotron emission results, which leads to very different fluxes. To agree with the upper limit of the radio surveys, the profiles require different annihilation cross-sections $\langle\sigma v\rangle$ once we had fixed the DM mass and all the other parameters are intrinsic to their own UCDs.

Using the set of parameters described in Table \ref{Tab1}, plus the parameters described in this section, an approximate solution which fits the radio spectrum with a DM candidate annihilating in $b\bar{b}$  with  $m_{DM}=34$ GeV or annihilating in $\mu^{+}\mu^{-}$  with  $m_{DM}=10$ GeV,  would require an annihilation cross-section as described in the Table \ref{Tab2}. We emphasize that these numbers can change once the diffusion effects are included in spiky profiles.

\begin{table}[h]
	\centering
	\begin{tabular}{ |p{3.7cm}||p{2.2cm}|p{2.0cm}|p{2.7cm}| p{2.2cm}| }
 \hline
 \multicolumn{5}{|c|}{UCD3} \\
 \hline
 & Fornax ($b\bar{b}$) & M59  ($b\bar{b}$) &Fornax ($\mu^{+}\mu^{-}$) &M59  ($\mu^{+}\mu^{-}$)\\
 \hline
 $ \langle\sigma v\rangle$ (Frank, $\rho=0.08$ GeV/cm$^{3}$)  & $3\times 10^{-26}$    & $3\times 10^{-26}$ &  $3\times 10^{-28}$ & $3\times 10^{-28}$\\\hline
  $ \langle\sigma v\rangle$ (Frank, $\rho=2.07$ GeV/cm$^{3}$)  & $3\times 10^{-26}$    & $3\times 10^{-26}$ &  $3\times 10^{-28}$ & $3\times 10^{-28}$\\\hline
  $ \langle\sigma v\rangle$ (NFW+Spike, $\rho=0.08$ GeV/cm$^{3}$) &   $1\times 10^{-35}$ &  $1\times 10^{-34}$  & $3\times 10^{-39}$ & $6\times 10^{-33}$\\\hline
  $ \langle\sigma v\rangle$ (NFW+Spike, $\rho=2.07$ GeV/cm$^{3}$) &   $1\times 10^{-41}$ &  $2\times 10^{-38}$  & $1\times 10^{-33}$ & $2\times 10^{-38}$\\\hline
  $ \langle\sigma v\rangle$ (Spike 32\%)  &$1\times 10^{-32}$  & $1\times 10^{-34}$ & $8\times 10^{-32}$ & $1\times 10^{-33}$\\\hline
  $ \langle\sigma v\rangle$ (Spike 8\%)  & $1\times 10^{-31}$  & $8\times 10^{-34}$ & $3\times 10^{-30}$ & $8\times 10^{-33}$ \\\hline
     \end{tabular}
    \caption{ Annihilation cross sections (cm$^{3}$/s) used in the synchrotron flux calculation considering Frank, NFW+Spike and Spike profiles. For Frank and NFW+Spike profile, we had considered the DM density $\rho=0.08$ GeV/cm$^{3}$ and $\rho=2.07$ GeV/cm$^{3}$. For the Spike profile we had considered that all the mass inside the spike corresponds to $\sim $ 32\% and/or 8\% of the UCD3 mass.}
    \label{Tab2}
\end{table}

In regard to the Frank profile,  for $DM,DM \rightarrow b\bar{b}$, we had taken the thermal cross section and for $DM,DM \rightarrow \mu^{+}\mu^{-}$, we had considered $ \langle\sigma v\rangle=3\times 10^{-28}$ cm$^{3}$/s. The use of this profile will produce fluxes well below experimental upper limits.

The other spike profiles that account for the interaction of DM \& SMBH have the power of increasing the resulting fluxes. Neglecting diffusion, in order to have a flux of the order of $\sim 7.8\mu$ Jy the NFW+Spike and the Spike profiles would need to have their annihilation cross-sections $\langle\sigma v\rangle$ decreased by many orders of magnitude in comparison to the thermal cross-section. We did not include the synchrotron fluxes results for the mini-spike profile in our graphs because we didn't find satisfactory physical solutions for $ \langle\sigma v\rangle$ in the $\rho_{sat}$ validity domain. Considering the range of parameters of the UCDs and our scenarios, the NFW+Spike profile was the profile that must empower the flux.

\begin{figure*}[ht!]
	\centering
	\begin{tabular}{cc}
		\includegraphics[width=0.5\linewidth]{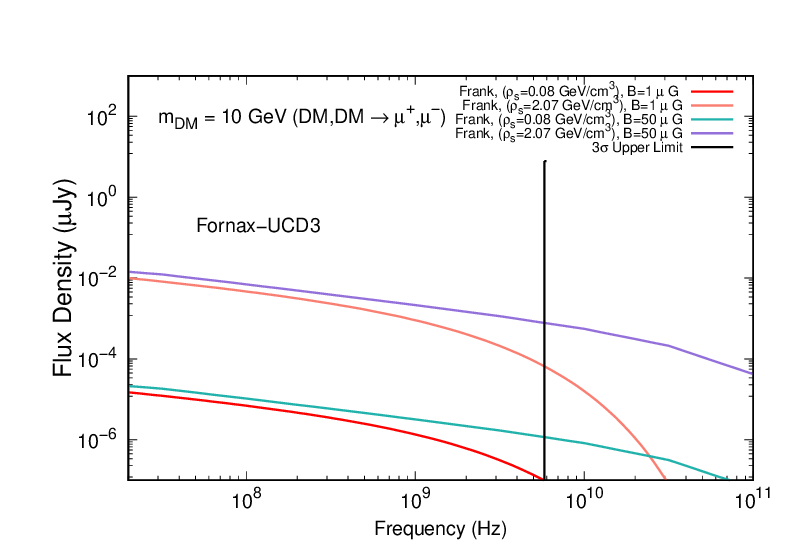}
		\includegraphics[width=0.5\linewidth]{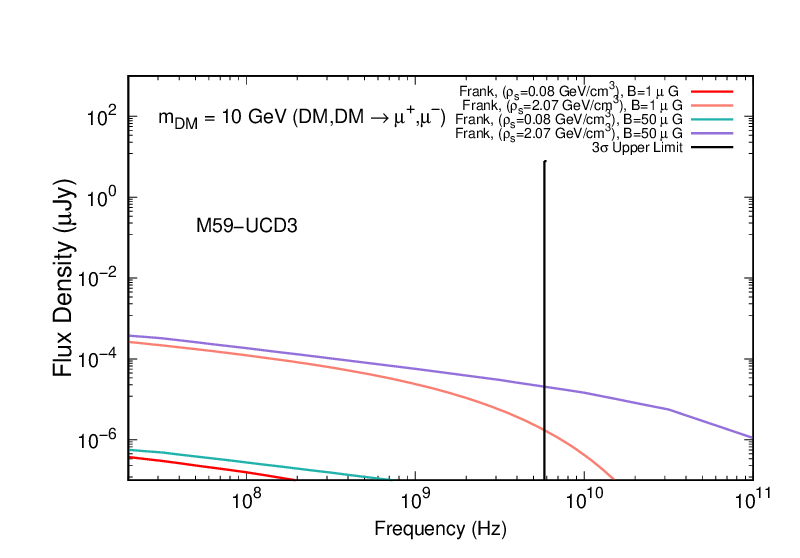}\\
		\includegraphics[width=0.5\linewidth]{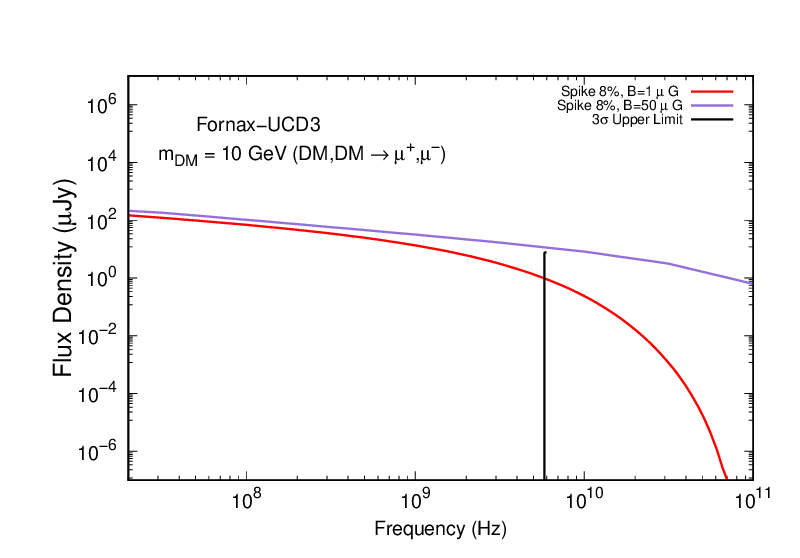}
		\includegraphics[width=0.5\linewidth]{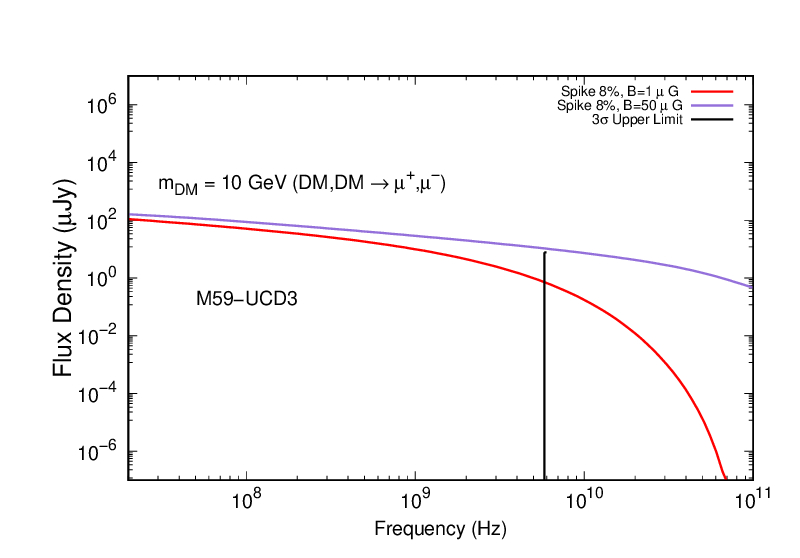}\\
		\includegraphics[width=0.5\linewidth]{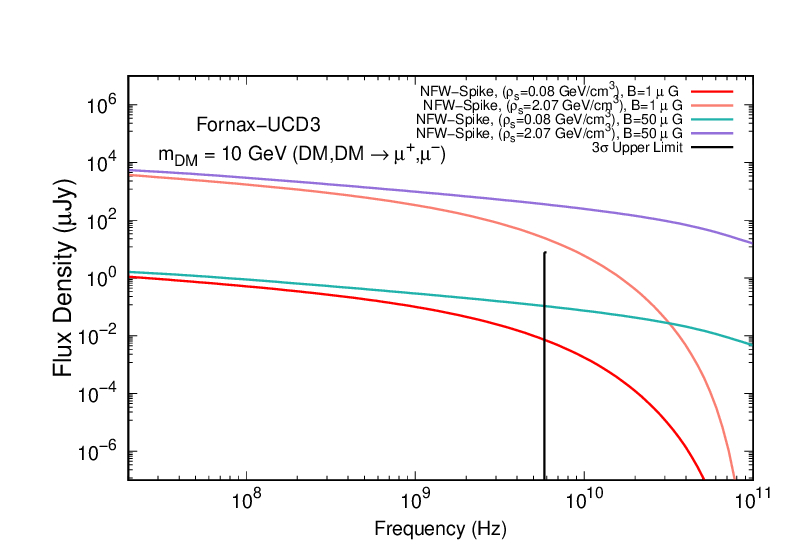}
		\includegraphics[width=0.5\linewidth]{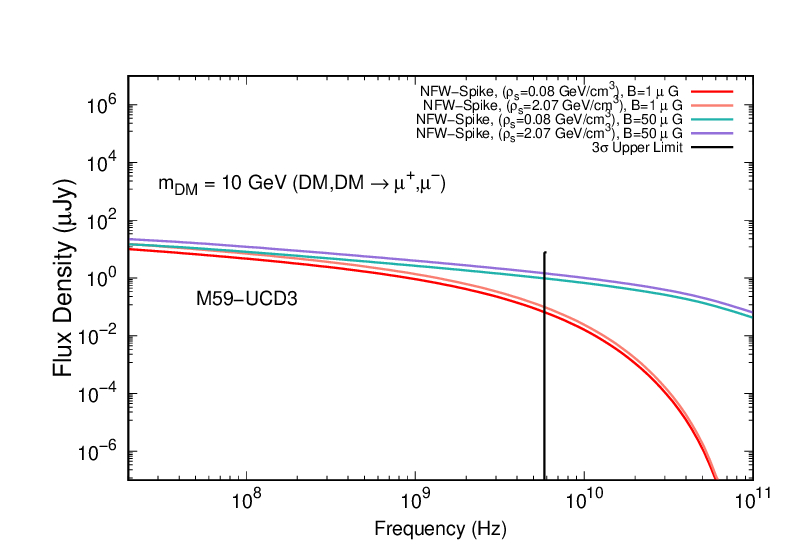}\\
	\end{tabular}
	\caption{The expected flux from DM annihilation as a function of frequency for Fornax UCD3 and M59 UCD3. In the upper panel, we show the synchrotron fluxes results for Fornax UCD3 and M59 UCD3 considering the Frank profile. In the middle panel, we present the synchrotron flux for Spike profile and in the lower panel we present the flux for the NFW+Spike profile.  Our DM candidate has  mass $m_{DM}=10$ GeV and annihilates in $\mu^{+}\mu^{-}$. All the values used for the annihilation cross-section can be found in Table \ref{Tab2}. M59 UCD3  has an experimental upper limit taken from deep radio imaging observations at 5.8 GHz \cite{Seth}, which is represented by the black straight line. The upper limit at that frequency is  7.8 $\mu$Jy. All these fluxes were built considering Frank, NFW+Spike, and Spike profiles with different DM density and DM concentration and we didn't consider the diffusion effect in this graphs.}
	\label{fig3}
\end{figure*}

In Figure \ref{fig3}, we show the synchrotron flux for Fornax UCD3 and M59 UCD3 considering the profiles Frank, Spike and  NFW+Spike with different magnetic field configuration, more specifically, $B=1\mu$G and $B=50\mu$G. These profiles are also presented with different critical density and/or concentration parameters. For $B=1\mu$G, the flux decreases faster for lower frequencies if compared to the flux built with $B=50\mu$G. The parameters used to build these graphs are the ones presented in Table \ref{Tab2}.

\newpage
\section{Conclusion and Future Perspectives}
\label{sec:Conclusion}
${}$

The formation mechanism of ultra-compact dwarf galaxies is still unknown. Although there is controversy in the literature about the presence or absence of dark matter in these systems, it seems more plausible that a certain fraction of DM exists at the level of galactic systems. DM plays a key role in building the large-scale structure of the Universe, from the galactic scale to the scale of galaxy clusters. Understanding the process of forming UCDs can, in large part, be connected to identifying how DM is distributed inside them or explaining why DM is absent from these systems.

In this work, for the first time, we have modeled the DM component in two UCDs: Fornax UCD3 and M59 UCD3. We have considered DM annihilating around an SMBH, using in our simulations some well-motivated spike profiles which enhance the presence of a  BH in these UCDs. We have explored the parameters which are most important to calculate the radio flux and have investigated some possible models of DM distribution focusing attention on a light DM candidate with masses 10 and 34 GeV, annihilating in  $\mu^{+}\mu^{-}$  and in $b\bar{b}$ respectively.

First we had studied the canonical NFW DM density profile. To build the synchrotron flux for this profile, we included the diffusion effects in our calculations. In fact, the diffusion is very important and decreases the flux by many orders of magnitude as compared to the fluxes with no diffusion effects.

In a future work we intend to pay much more attention to the diffusion effects in spiky density profiles especially studying the strong enhancements of such profiles considering  the dynamical relaxation by stellar interactions, the loss of DM falling into the BH, and other dynamical effects of BH-DM interaction.

With our simulations for spiky profiles, neglecting diffusion, we conclude that, the size of the spike and the internal properties of the black hole at these UCDs can leave their imprint in synchrotron flux. This imprint can be very strong as compared to those predicted from simulations of Fornax UCDs \cite{Frank:2011ji}. The DM density distribution that mimics DM in Fornax UCD3 \cite{Frank:2011ji} produces synchrotron fluxes well below the upper limits predicted for M59 UCD3 when we use annihilation cross-sections $\langle\sigma v\rangle$ in the same order or two orders of magnitude below the thermal one. For the Spike profiles, considering the set of parameters studied here, the annihilation cross-sections should be many orders of magnitude below the thermal cross-section to agree with the upper limits for these UCDs, but this conclusion may change considering the diffusion and other strong enhancements of BH-DM interaction.

Motivated by the morphological characteristics of Fornax UCD3 and M59 UCD3, to compare our synchrotron results we have taken the upper limit for radio emission predicted for M59 UCD3 and applied it also for Fornax UCD3.  We conclude that future radio surveys observing Fornax UCD3 have the potential to reveal or constrain some signal of DM annihilation. In the absence of a gamma-ray signal, synchrotron emission is an interesting method for testing the potential astrophysical signature of DM, especially in the regions where DM and the black holes might be present.

${}$

\section*{Acknowledgments}
 E. C. F. S. Fortes and O. D. Miranda thanks FAPESP for financial support under contract 2018/21532-4.

\end{document}